\documentclass[journal=jctcce,manuscript=article]{achemso}

\usepackage[version=3]{mhchem} 

\usepackage{amsmath}
\usepackage{graphicx,color}
\usepackage{amsfonts}
\usepackage{algorithm,algorithmic}
\usepackage{epstopdf}
\usepackage{comment}

\usepackage[version=3]{mhchem} 

\newcommand{\abs}[1]{\left\lvert#1\right\rvert}
\newcommand{\norm}[1]{\left\lVert#1\right\rVert}

\newcommand{\ud}{\,\mathrm{d}}

\DeclareMathOperator{\Tr}{Tr}

\newcommand{\Vion}{V_{\mathrm{ion}}}
\newcommand{\onlinecite}[1]{\citenum{#1}}
\newcommand{\bvec}[1]{\mathbf{#1}}
\newcommand{\vr}{\bvec{r}}
\newcommand{\vR}{\bvec{R}}

\newcommand{\CY}[1]{{\color{red}~\textsf{[CY: #1]}}}

\author{Wei Hu}
\email{whu@lbl.gov}
\affiliation{Computational Research Division,
Lawrence Berkeley National Laboratory, Berkeley, California 94720,
United States}

\author{Lin Lin}
\email{linlin@math.berkeley.edu}
\affiliation{Department of
Mathematics, University of California, Berkeley, California 94720,
United States} \alsoaffiliation{Computational Research Division,
Lawrence Berkeley National Laboratory, Berkeley, California 94720,
United States}

\author{Chao Yang}
\email{cyang@lbl.gov}
\affiliation{Computational Research Division,
Lawrence Berkeley National Laboratory, Berkeley, California 94720,
United States}

\title[PC-DIIS]{Projected Commutator DIIS Method for Accelerating Hybrid
Functional Electronic Structure Calculations}

\begin{document}

\begin{abstract}
The commutator direct inversion of the iterative subspace
(commutator DIIS or C-DIIS) method developed by Pulay is an
efficient and the most widely used scheme in quantum chemistry to
accelerate the convergence of self consistent field (SCF) iterations
in Hartree-Fock theory and Kohn-Sham density functional theory. The
C-DIIS method requires the explicit storage of the density matrix,
the Fock matrix and the commutator matrix. Hence the method can only
be used for systems with a relatively small basis set, such as the
Gaussian basis set. We develop a new method that enables the C-DIIS
method to be efficiently employed in electronic structure
calculations with a large basis set such as planewaves for the first
time. The key ingredient is the projection of both the density matrix and
the commutator matrix to an auxiliary matrix called the gauge-fixing
matrix. The resulting projected commutator-DIIS method (PC-DIIS)
only operates on matrices of the same dimension as the that 
consists of Kohn-Sham orbitals. The cost of the method is comparable to that of
standard charge mixing schemes used in large basis set calculations.
The PC-DIIS method is gauge-invariant, which guarantees that its
performance is invariant with respect to any unitary transformation of the Kohn-Sham
orbitals. We demonstrate that the PC-DIIS method can be viewed as an
extension of an iterative eigensolver for nonlinear problems.  We
use the PC-DIIS method for accelerating Kohn-Sham density functional
theory calculations with hybrid exchange-correlation functionals,
and demonstrate its superior performance compared to the commonly
used nested two-level SCF iteration procedure.  Furthermore, we
demonstrate that in the context of \textit{ab initio} molecular
dynamics (MD) simulation with hybrid functionals, one can
extrapolate the gauge-fixing matrix to achieve the goal of
extrapolating the entire density matrix implicitly along the MD
trajectory. Numerical results indicate that the new method
significantly reduces the number of SCF iterations per MD step,
compared to the commonly used strategy of extrapolating the electron
density.
\end{abstract}

\section{Introduction} \label{sec:Introduction}

In electronic structure calculations based on the Hartree-Fock (HF)
theory and the Kohn-Sham density functional theory (KSDFT), the
density matrix needs to be computed self-consistently. For quantum
chemistry software packages based on Gaussian orbitals or localized
atomic orbitals, the most widely used numerical scheme to achieve
self-consistency is the commutator direct inversion of the iterative
subspace (commutator DIIS or C-DIIS)
method.\cite{JCC_3_54_1982_DIIS} Although there is no known
theoretical guarantee for the convergence of the C-DIIS method,
numerous numerical results indicate that the method often converges
rapidly starting from almost any initial guesses. However, the
C-DIIS method requires the explicit storage of the density matrix,
the Fock matrix and the commutator matrix. Hence the method is
limited to systems discretized using a small basis set such as
Gaussian orbitals. For large basis sets such as planewaves, it is
prohibitively expensive to store the full Fock matrix or the
density matrix, and the C-DIIS method is not directly applicable.
Instead, in large basis set calculations, the most commonly used
method is to only perform mixing on local quantities such as the
electron density or the local potential, which corresponds to the
diagonal elements of the density matrix and the Fock matrix in the
real space representation, respectively. When combined with a good
preconditioner,\cite{PRB_23_3082_1981_Kerker,PRB_54_11169_1996_RMM-DIIS}
the density mixing and potential mixing schemes can also be highly
efficient for KSDFT calculations with local and semi-local
exchange-correlation functionals, such as the local density
approximation (LDA),\cite{PRL_45_566_1980, PRB_23_5048_1981,
PRB_54_1703_1996_LDA} the generalized gradient approximation
(GGA),\cite{PRA_38_3098_1988, PRB_37_785_1988, PRL_77_3865_1996_PBE}
and meta-GGA functionals.\cite{PRL_91_146401_2003_TPSS,
PRL_115_036402_2015_SCAN, PNAS_112_685_2015}

Hybrid exchange-correlation functionals, such as
B3LYP,\cite{JCP_98_1372_1993} PBE0\cite{JCP_105_9982_1996} and
HSE\cite{JCP_118_8207_2003, JCP_124_219906_2006_HSE06}, are known to be
more reliable in producing high fidelity results for electronic
structure calculations. Hybrid functionals include a fraction of the
Fock exchange operator, which not only depends on the electron density
but also the off-diagonal elements of the density matrix.  This leads to
significant increase of the computational cost compared to calculations
with semi-local functionals. For calculations performed in large basis sets
such as planewaves, an iterative diagonalization procedure is used solve
the Kohn-Sham equations. The diagonalization procedure requires 
multiplying the exchange operator with the occupied orbitals in each 
iteration.  These multiplications alone often constitutes
more than $95\%$ of the overall computational time in a conventional
approach. Many numerical methods have been developed to reduce the cost of such multiplication operations.
Notably, linear scaling methods \cite{RMP_71_1085_1999_ON,
RPP_75_036503_2010_ON, JCTC_6_2348_2010, JCP_141_084502_2014,
JCTC_11_4655_2015, JCTC_11_1463_2015} construct a sparse approximation
to the exchange operator, and have been developed and applied to large
systems with substantial band gaps. Recently, we have developed the adaptively
compressed exchange operator (ACE)
formulation,~\cite{JCTC_12_2242_2016_ACE,JCTC_13_1188_2017_ACE} and
the interpolative separable density fitting  (ISDF)
method\cite{LuYing2015,HuLinYang2017} to reduce the cost associated with
the exchange operator. Unlike linear scaling methods, one notable
feature of ACE and ISDF is that these methods are insensitive to
the energy band gap of the target system, and hence can be applied to
insulating, semiconducting, and even metallic systems.

In this paper, we are concerned with yet another difficulty in hybrid
functional calculations related to the SCF iterations, which is
somewhat orthogonal to the difficulty introduced by the large
computational cost associated with the multiplication of the exchange
operator. Note that
self-consistency must be achieved for the entire density matrix. For
algorithms that use a large basis set 
to discretize the problem, it is not practical to store and update
the density matrix directly. The commonly used approach to achieve
self-consistency is a nested two-level SCF iteration procedure that
consists of an inner SCF loop and an outer SCF loop. In the inner
SCF loop, the density matrix and hence the Fock exchange operator
are fixed (implicitly by fixing the Kohn-Sham orbitals), and the SCF
iteration is only performed for the charge density. Once the error
of the inner iteration reaches below certain tolerance level, the
density matrix is updated using a fixed point iteration in the outer
iteration, which is carried out implicitly through a fixed point
iteration for all the Kohn-Sham orbitals.  The use of the two nested
SCF loops allows charge mixing schemes such as Anderson
mixing\cite{JACM_12_547_1965_Anderson} and Kerker
preconditioner\cite{PRB_23_3082_1981_Kerker} to be performed to
avoid the ``charge sloshing''
phenomenon.\cite{PRB_54_11169_1996_RMM-DIIS,AriasPayneJoannopoulos1992}
However, it also significantly increases the number of iterations
for hybrid functional calculations to converge. 
There are other
methods for accelerating the SCF iteration, such as the E-DIIS
method.\cite{KudinScuseriaCances2002} Besides the SCF iteration
procedure, HF and KSDFT calculations can also be performed through
direct minimization of the orbitals without storing the density
matrix, such as the direct optimization
method\cite{AriasPayneJoannopoulos1992,YangMezaWang2007,WenYin2010}
and the Car-Parrinello method\cite{CarParrinello1985} with a damped
dynamics. The advantage of these methods is that one can use the
same algorithmic structure for semi-local and hybrid functional
calculations. However, the convergence rate of these methods may be
slower when compared to SCF iteration methods such as C-DIIS.

In this paper, we develop a new method that enables the C-DIIS
method to be efficiently employed in electronic structure
calculations with a large basis set for the first time. As a result,
the two level SCF loops can be efficiently reduced into a single
loop, which significantly cuts down the total number of SCF
iterations.  Our method never requires storing the density matrix,
Fock matrix or the commutator. It operates on matrices with $O(N_e)$ 
columns, where $N_e$ is the number of electrons.  The key idea is
to project both the density matrix and the commutator to an
auxiliary matrix called the gauge-fixing matrix. The resulting
projected commutator DIIS method (PC-DIIS) is inherently
gauge-invariant and hence numerically stable. We demonstrate that
the PC-DIIS method can be viewed as an extension of an iterative
eigensolver for nonlinear problems. The PC-DIIS method is well
suited for accelerating Hartree-Fock and hybrid functional Kohn-Sham
density functional theory calculations. The PC-DIIS method can also
be naturally combined with other recently developed techniques for
accelerating hybrid functional calculations such as ACE and ISDF.
Similar to ACE and ISDF, we find that the
effectiveness of the PC-DIIS method is not sensitive to the size of the band gap.


In the context of \textit{ab initio} molecular dynamics (AIMD)
simulation, since it is not practical to store the entire density
matrix, one commonly used strategy is to only extrapolate the
electron density from one time step to another. This leads to a
mismatch between the electron density and the density matrix defined
by the Kohn-Sham orbitals. We demonstrate that in the PC-DIIS
method, one can extrapolate the gauge-fixing matrix along the MD
trajectory. The gauge-fixing matrix can be chosen to be smooth with
respect to the time $t$, and to carry equivalent information as in
the density matrix at each time step. As a result, the density
matrix is implicitly extrapolated along the MD trajectory. This
procedure shares similarity with the predictor-corrector scheme for
extrapolating Kohn-Sham orbitals~\cite{KuehneKrackMohamedEtAl2007}.
Compared to the method that only extrapolates the electron density,
we find that extrapolating the gauge-fixing matrix can lead to
significant reduction of the number of SCF iterations per MD step.

We demonstrate the performance of the PC-DIIS method for a number of
systems with insulating and metallic characters. We find that the
PC-DIIS method can significantly reduce the number of iterations and
hence the wall clock time. For instance, using the HSE06 hybrid
functional, we can fully converge a bulk silicon system with $1000$
atoms within $5$ minutes of wall clock time.

The rest of the manuscript is organized as follows. We introduce
the two level nested SCF method and the C-DIIS method in
section~\ref{sec:nested} and~\ref{sec:cdiis}, respectively. We present
the new PC-DIIS method in section~\ref{sec:pcdiis}, and demonstrate how
to extrapolate the gauge-fixing matrix in the context of AIMD simulation in
section~\ref{sec:aimd}. Numerical results are presented in
section~\ref{sec:Result}, followed by a conclusion and discussion in
section~\ref{sec:conclusion}.

\section{Two level nested SCF method}\label{sec:nested}

The Hartree-Fock-like equations as appeared in the Hartree-Fock theory
and the Kohn-Sham density functional theory with hybrid functionals are a
set of nonlinear equations as follows
\begin{equation}
  \begin{split}
    &H[P]\psi_{i} = \left(-\frac12 \Delta  +
    \Vion +
    V_{\text{Hxc}}[P] + V_{X}[P]\right)\psi_{i} = \varepsilon_{i} {\psi}_{i},\\
    &\int {\psi}^{*}_{i}(\vr) {\psi}_{j}(\vr) \ud \vr = \delta_{ij},
    \quad
    P(\vr,\vr') = \sum_{i=1}^{N_e} \psi_i(\vr) \psi_i^*(\vr').
  \end{split}
  \label{eqn:HF}
\end{equation}
Here $H[P]$ is called a Hamiltonian operator (also called a Fock
operator). The eigenvalues $\{\varepsilon_{i}\}$ are ordered
non-decreasingly, and $N_e$ is the number of electrons (spin
degeneracy omitted for simplicity). $P$ is the density matrix
associated with $\{\psi_i\}$, $i=1,2...,N_e$. It is an orthogonal
projector with an exact rank $N_e$. The diagonal entries of $P$ give
the electron density $\rho(\vr)=P(\vr,\vr)$. $\Vion$ characterizes
the electron-ion interaction in all-electron calculations.
$V_{\text{Hxc}}$ is a local operator that characterizes the Hartree
and the exchange-correlation contributions modeled at a local or
semi-local level. It typically depends only on the electron density
$\rho(\vr)$. The exchange operator $V_{X}$ is an integral operator
with a kernel
\begin{equation}
  V_{X}[P](\vr,\vr') =
  -P(\vr,\vr') K(\vr,\vr'),
  \label{eqn:VXkernel}
\end{equation}
where $K(\vr,\vr')$ is an operator that accounts for the
electron-electron interaction. For example, in the Hartree-Fock
theory, $K(\vr,\vr')=1/\abs{\vr-\vr'}$ is the Coulomb operator. In
screened exchange theories such as HSE\cite{JCP_118_8207_2003,
JCP_124_219906_2006_HSE06}, $K$ is a screened Coulomb operator
$K(\vr,\vr')=\text{erfc}(\mu\abs{\vr-\vr'})/\abs{\vr-\vr'}$.

Methods for solving the Hartree-Fock equation often use an iterative
procedure (i.e.,the SCF iteration) in which the density matrix $P$
is updated until it is consistent with the Hamiltonian operator
$H[P]$. When a large basis set such as the planewave basis set is
used, it becomes prohibitively expensive to store the density matrix
$P$. On the other hand, one cannot simply take the output Kohn-Sham
orbitals from one SCF iteration and use them as the input Kohn-Sham
orbitals for the next SCF iteration. This type of fixed point
iteration is known to suffer from the ``charge sloshing''
problem.\cite{PRB_54_11169_1996_RMM-DIIS,AriasPayneJoannopoulos1992}
In practice, the most commonly used method, such as the one
implemented in the Quantum ESPRESSO software
package,\cite{JPCM_21_395502_2009_QE} uses a two-level nested SCF
procedure.  The motivation for using a two-level SCF procedure is to
apply advanced charge mixing schemes to the electron density $\rho$
in the inner iteration to mitigate charge sloshing, and to use a
fixed point iteration to update the Kohn-Sham orbitals and
consequently the exchange potential in the outer iteration. The
update of the exchange potential is more costly, even though its
contribution to total energy is typically much smaller.

The two-level nested SCF method is summarized in
Alg.~\ref{alg:twoloop}. In each outer iteration, the exchange
operator $V_{X}[P]$ is updated. This is implicitly done by updating
a set of orbitals $\{\varphi_{i}\}_{i=1}^{N_{e}}$ defining the
density matrix as $P=\sum_{i=1}^{N_{e}}\varphi_{i}\varphi_{i}^{*}$.
We remark that this set of orbitals may be different from the
Kohn-Sham orbitals in the inner SCF iteration. The update is done
through a fixed point iteration, i.e.
$\{\varphi_{i}\}_{i=1}^{N_{e}}$ are given by the output Kohn-Sham
orbitals in the previous outer iteration.  In the inner SCF
iteration, with the exchange operator fixed, the Hamiltonian $H$
only depends on the electron density $\rho$. Charge mixing schemes
for $\rho$ can be performed in similar fashion to what is done in 
a standard KSDFT
calculation without the exchange operator in the inner SCF
iteration. Finally, within each inner iteration, with both $P$ and
$\rho$ fixed, Eq.~\eqref{eqn:HF} becomes a linear eigenvalue problem
and can be solved by an iterative eigensolver such as the Davidson
method,\cite{JCP_17_87_1975_Davidson} the LOBPCG
method,\cite{SIAMJSC_23_517_2001_LOBPCG} or the PPCG
method.\cite{JCP_290_73_2015_PPCG}  The outer SCF iteration
continues until convergence is reached, which can be monitored e.g.
in terms of the change of the exchange energy.

\begin{algorithm}[H]
  \caption{Two-level nested SCF method for solving Hartree-Fock-like equations.}

  \begin{algorithmic}[1]

    \WHILE {Exchange energy is not converged}

    \WHILE {Electron density $\rho$ is not converged}

    \STATE Solve the linear eigenvalue problem $H\psi_{i} =
    \varepsilon_{i} {\psi}_{i}$ with any iterative eigensolver.

    \STATE Update $\rho^\text{out}(\mathbf{r})\gets \sum_{i=1}^{N_e}
    \abs{{\psi}_{i}(\mathbf{r})}^2$.

    \STATE Update $\rho$ using $\rho^\text{out}$ and charge densities computed and saved from the previous iteration using a charge mixing scheme.

  \ENDWHILE

  \STATE Compute the exchange energy.

  \STATE Update $\{\varphi_{i}\}_{i=1}^{N_{e}}\gets
  \{\psi_{i}\}_{i=1}^{N_{e}}$.
\ENDWHILE

 \end{algorithmic}
 \label{alg:twoloop}
 \end{algorithm}

\section{Commutator DIIS method}\label{sec:cdiis}


In the two-level nested SCF procedure, the density matrix is only
updated implicitly in the outer iteration by a fixed point iteration
of the Kohn-Sham orbitals. The overall convergence of this method is
slower than an alternative scheme in which the density matrix and
the Fock matrix are updated, e.g. by the commutator direct inversion
of the iterative subspace method (commutator DIIS or C-DIIS)
developed by Pulay.\cite{JCC_3_54_1982_DIIS} The C-DIIS method,
which requires saving density and Fock matrices for a few
iterations, is often used in many hybrid DFT and Hartree-Fock
calculations performed in a small basis sets. However, because it is
too costly to compute and save the entire density and Fock matrices
explicitly in each SCF for a calculation performed in a large basis
set, the direct use of the C-DIIS method is not feasible.

Before we introduce an efficient way to perform C-DIIS in a hybrid
DFT or Hartree-Fock calculation performed in a large basis, we first
briefly introduce the standard C-DIIS method below. For simplicity,
we assume the Kohn-Sham or Hartree-Fock equations are discretized by
an orthonormal basis set. We define the residual $R$ as the
commutator between $H[P]$ and $P$, i.e.,
\begin{equation}
  R[P] = H[P] P - P H[P].
  \label{eqn:residual}
\end{equation}
The C-DIIS method is designed to minimize the residual within a
subspace that contains the residuals associated with previous
approximations to the Fock and density matrices. To be specific, if
$H^{(k)}$ denotes the approximate Hamiltonian produced at step $k$,
we define a new Hamiltonian $\tilde{H}^{(k+1)}$ at step $(k+1)$ as a
linear combination of $\ell$ previous approximations to the
Hamiltonian, i.e.,
\begin{equation}
  \tilde{H}^{(k+1)} = \sum_{j=k-\ell}^{k}\alpha_{j} H^{(j)},
  \label{eqn:Hkp1}
\end{equation}
where $\alpha_{j}$ satisfies the constraint $\sum_{j=k-\ell}^{k}
\alpha_{j}=1$. Each Hamiltonian $H^{(j)}$ defines a density matrix
$P^{(j)}$ through the solution of the linear eigenvalue
problem~\eqref{eqn:HF}. Before self-consistency is reached, the
residual $R^{(j)}=R[P^{(j)}]$ defined by Eq.~\eqref{eqn:residual} is
nonzero. However, when all the Hamiltonian matrices $\{H^{(j)}\}$
are close to the self-consistent Hamiltonian operator $H^{\star}$,
it is reasonable to expect the residual associated with
$\tilde{H}^{(k+1)}$ to be well approximated by $\tilde{R}^{(k+1)}
\equiv \sum_{j=k-\ell+1}^{k}\alpha_{j} R^{(j)}$. The C-DIIS method
determines $\{\alpha_{j}\}$ by minimizing $\tilde{R}^{(k+1)}$ with
respect to $\alpha_j$'s, i.e., we solve the following constrained
minimization problem in each C-DIIS iteration.
\begin{equation}
    \min_{\{\alpha_{j}\}} \norm{\sum_{j=k-\ell}^{k}\alpha_{j}
    R^{(j)}}^2_{F}, \qquad
    \text{s.t.} \quad \sum_{j=k-\ell}^{k}\alpha_{j} = 1.
  \label{eqn:cdiiscon}
\end{equation}
Here the Frobenius norm is defined as $\norm{A}_{F}^2=\Tr[A^{*}A]$.
Although a vanishing commutator residual is only a necessary
condition for reaching self-consistency, it has been found that, in
the context of Hartree-Fock equation, the commutator corresponds to
the gradient direction of the total energy with respect to $P$.
Hence the C-DIIS method can be interpreted as a quasi-Newton
method.\cite{KudinScuseria2007}

Note that the constraint in~\eqref{eqn:cdiiscon} can be eliminated by
rewriting Eq.~\eqref{eqn:Hkp1} as
\begin{equation}
  \tilde{H}^{(k+1)} = H^{(k)} + \sum_{j=k-\ell+1}^{k}\beta_{j}
  (H^{(j-1)} - H^{(j)}).
  \label{eqn:Hkp2}
\end{equation}
The relation between Eq.~\eqref{eqn:Hkp1} and~\eqref{eqn:Hkp2} can be
seen from the following mapping between $\{\alpha_i\}$ and
$\{\beta_j\}$ parameters, i.e.,
\begin{equation}
  \alpha_{j} = \begin{cases}
    1-\beta_{j}, & j=k,\\
    \beta_{j+1}-\beta_{j}, & k-\ell+1\le j\le k-1, \\
    \beta_{j+1}, & j=k-\ell.
  \end{cases}
  \label{}
\end{equation}
Clearly
$\sum_{j=k-\ell}^{k}\alpha_{j} = 1$ is satisfied, and the new variables
$\beta=(\beta_{k-\ell+1},\ldots,\beta_{k})^{T}$
become unconstrained. Define
$Y^{(j)}=R^{(j-1)}-R^{(j)}$ for $k-\ell+1\le j\le k$, then the
constraint minimization problem~\eqref{eqn:cdiiscon} becomes a
unconstrained minimization problem
\begin{equation}
  \min_{\{\beta_{j}\}} \norm{R^{(k)}+\sum_{j=k-\ell+1}^{k}\beta_{j}
  Y^{(j)}}^2_{F}.
  \label{eqn:cdiisuncon}
\end{equation}
As a result, Eq.~\eqref{eqn:cdiisuncon} has an analytic solution
\begin{equation}
  \beta = -M^{-1} b,
  \label{eqn:betasol}
\end{equation}
where the $\ell\times \ell$ matrix $M$ and the vector $b$ are
defined as
\begin{equation}
  M_{ij} = \Tr[(Y^{(i)})^{*}Y^{(j)}], \quad b_{j} =
  \Tr[(Y^{(j)})^{*}R^{(k)}]
  \label{}
\end{equation}
respectively.

Using the solution~\eqref{eqn:betasol}, we obtain
$\tilde{H}^{(k+1)}$ via~\eqref{eqn:Hkp2}. This procedure is repeated
until the residual $R[P]$ is sufficiently small. We remark that the
solution in the form of~\eqref{eqn:betasol} can also be interpreted
as a variant of the Anderson acceleration
method.\cite{JACM_12_547_1965_Anderson}


\section{Projected commutator DIIS}\label{sec:pcdiis}

In this section, we introduce a new method that enables the C-DIIS
method to be used for hybrid DFT or Hartree-Fock calculations
performed within a large basis set. Since it is not possible to
explicitly store or mix the density matrices in such calculations,
it is tempting to perform the DIIS procedure on the $N_{g}\times
N_{e}$ orbital matrix $\Psi=[\psi_{1},\ldots,\psi_{N_{e}}]$, where
$N_{g}$ is the number of degrees of freedom to discretize each
orbital (e.g. the number of planewaves), and $N_{e}$ is the number
of electrons or Kohn-Sham orbitals. However, one key difference
between the density matrix and the orbital matrix is that the former
is gauge invariant. That is, if we replace $\Psi$ by $\Psi U$ where
$U$ is an $N_{e}\times N_{e}$ unitary gauge matrix, the density
matrix $P=\Psi\Psi^*=\Psi U U^* \Psi^*$ does not change. Therefore,
it is completely safe to combine two density matrices constructed
from $\Psi$'s that differ by a gauge transformation. It is important
to note that in both the Hartree-Fock theory and KSDFT with
semi-local and hybrid functionals, the total energy is invariant
with respect to any gauge transformation.

However, since the orbital matrix $\Psi$ is not gauge invariant,
combining successive approximations to $\Psi$'s that differ by a
gauge transformation may hinder the convergence of the SCF
iteration, or may not be stable at all.  This type of scenario may
arise when eigenvalues of $H$ are degenerate or nearly degenerate. A
slight change in the potential in $H$ from one SCF iteration to
another may lead to an arbitrary rotation of the eigenvectors
associated with these eigenvalues.\cite{GolubVan2013} Consequently,
a linear combination of the orbital matrices associated with these
two consecutive SCF iterations is not expected to be effective.

To overcome this difficulty, we introduce an auxiliary orbital
matrix $\Phi$ that spans the same subspace spanned by $\Psi$. This
orbital matrix is obtained by applying the orthogonal projection
operator associated with $\Psi$ to a reference orbital matrix
$\Phi_{\text{ref}}$ to be specified later in this section.  That is,
$\Phi$ is chosen to be
\begin{equation}
  \Phi = P \Phi_{\text{ref}} = \Psi (\Psi^{*}\Phi_{\text{ref}}).
  \label{eqn:Phitransform}
\end{equation}
We require $\Phi_{\text{ref}}$ to be fixed throughout the entire SCF
procedure. Note that $\Phi$ is invariant to any gauge transformation
applied to $\Psi$. Therefore, the auxiliary orbital matrices
obtained in successive SCF iterations can be safely combined to
produce a better approximation to the desired invariant subspace.



The columns of $\Phi$ are generally not orthogonal to each other.
However, as long as the columns of $\Phi$ are not linearly
dependent, both $\Psi$ and $\Phi$ span the range of the density
matrix $P$, which can also be written as
\begin{equation}
  P = \Phi (\Phi^{*}\Phi)^{-1} \Phi^{*}.
  \label{eqn:PPhi}
\end{equation}

The new method we propose to accelerate the SCF iteration for hybrid
DFT and HF calculations performed in a large basis set, which we
call a projector commutator DIIS (PC-DIIS) method, constructs a new
approximation to $\Phi$ in the $k$th SCF iteration by taking a
linear combination of the auxiliary orbital matrices
$\Phi^{(k-\ell)},\ldots,\Phi^{(k)}$ obtained in $\ell+1$ previous
iterations, i.e.
\begin{equation}
  \tilde{\Phi}^{(k+1)} = \sum_{j=k-\ell}^{k} \alpha_{j} \Phi^{(j)}.
  \label{eq:combphi}
\end{equation}
The coefficients $\{\alpha_{j}\}$ in \eqref{eq:combphi} are
determined by minimizing the residual associated with
$\tilde{\Phi}^{(k+1)}$, which, under the assumption that
$\Phi^{(j)}$ are sufficiently close to the solution of the Kohn-Sham
equations, is well approximated by $R \equiv
\sum_{j=k-\ell}^{k}\alpha_{j} R_{\Phi^{(j)}}$, where the residual
associated with an auxiliary orbital matrix $\Phi$ is defined by
\begin{equation}
  R_{\Phi} = H[P] P \Phi_{\text{ref}} - PH[P] \Phi_{\text{ref}}
  = (H[P]\Psi) \left(\Psi^{*}\Phi_{\text{ref}}\right) - \Psi
  \left((H[P]\Psi)^{*}\Phi_{\text{ref}}\right).
  \label{eqn:residualC}
\end{equation}

Note that evaluation of the residual in Eq.~\eqref{eqn:residualC}
only requires multiplying $H[P]$ with $\Psi$ and the multiplications
of matrices of sizes $N_{g}\times N_{e}$ and $N_{e}\times N_{e}$
only. These operations are already used in iterative methods for
computing the desired eigenvectors $\Psi$ of $H$. The PC-DIIS
algorithm does not require $P,H[P]$ or $R$ to be constructed or
stored explicitly.

An interesting observation is that, if $\Phi_{\mathrm{ref}}=\Psi$,
then $\Psi^{*}H[P]\Psi$ is a diagonal matrix denoted by $\Lambda$.
Consequently, the projected commutator takes the form
\begin{equation}
  R_{\Phi} = H[P] \Psi - \Psi \Lambda.
  \label{}
\end{equation}
This expression coincides with the standard definition of the
residual associated with an approximate eigenpair $(\Lambda,\Psi)$.
Hence the PC-DIIS method can also be viewed as an extension of an
iterative eigensolver for nonlinear problems.

Similar to the reformulation of the constrained minimization problem
into an unconstrained minimization problem in the C-DIIS method
presented in the previous section, the constrained minimization
problem
\begin{equation}
    \min_{\{\alpha_{j}\}} \norm{\sum_{j=k-\ell}^{k}\alpha_{j}
    R_{\Phi}^{(j)}}^2_{F}, \qquad
    \text{s.t.} \quad \sum_{j=k-\ell}^{k}\alpha_{j} = 1
  \label{eqn:pcdiiscon}
\end{equation}
to be solved in the PC-DIIS method can also be reformulated as an
unconstrained minimization problem. Using the same change of
variable as that presented in section~\ref{sec:cdiis}, we can write
\begin{equation}
  \tilde{\Phi}^{(k+1)} = \Phi^{(k)} + \sum_{j=k-\ell+1}^{k} \beta_{j}
  (\Phi^{(j-1)}-\Phi^{(j)}).
  \label{eqn:Pcp2}
\end{equation}
If we let $Y_{\Phi^{(j)}}=R_{\Phi^{(j-1)}}-R_{\Phi^{(j)}}$,
the coefficients $\beta_j$'s in \eqref{eqn:Pcp2} can be retrieved from
the vector $\beta=-(M^{\Phi})^{-1} b^{\Phi}$, where
\begin{equation}
  M^{\Phi}_{ij} = \Tr\left[(Y_{\Phi^{(i)}})^{*}Y_{\Phi^{(j)}}\right], \quad b^{\Phi}_{j} =
  \Tr\left[(Y_{\Phi^{(j)}})^{*}R_{\Phi^{(k)}}\right].
  \label{}
\end{equation}
Once $\tilde{\Phi}^{(k+1)}$ is obtained, a density matrix associated
with this orbital matrix is implicitly defined through
Eq.~\eqref{eqn:PPhi}. This implicitly defined density matrix allows
us to construct a new Hamiltonian from which a new set of Kohn-Sham
orbitals $\Psi^{(k+1)}$ and auxiliary orbitals $\Phi^{(k+1)}$ can be
computed.

We now discuss how to choose the gauge-fixing matrix
$\Phi_{\mathrm{ref}}$. Note that in hybrid functional calculations,
the contribution from the exchange operator is relatively small.
Hence the density matrix associated with Kohn-Sham orbitals obtained
from a DFT calculation that uses a local or semi-local
exchange-correlation functional is already a good initial guess
for the density matrix required in a hybrid functional calculation.
Therefore, we may use these orbitals as $\Phi_{\mathrm{ref}}$.
Compared to the two-level nested loop structure, the PC-DIIS method
only requires one level of SCF iteration. The PC-DIIS method is
summarized in Algorithm~\ref{alg:pcdiis}.

\begin{algorithm}[H]
  \caption{The PC-DIIS method for solving Hartree-Fock-like equations.}

  \begin{minipage}{\textwidth}
\begin{tabular}{p{0.5in}p{4.5in}}
{\bf Input}:  &  \begin{minipage}[t]{4.0in}
                 Reference orbitals $\Phi_{\mathrm{ref}}$.
                  \end{minipage} \\
{\bf Output}:  &  \begin{minipage}[t]{4.0in}
                 Approximate solution $\Psi = \{\psi_i\}$, $i=1,2...,N_e$,
                 to Eq.~\eqref{eqn:HF}.
                  \end{minipage}
\end{tabular}
\end{minipage}
  \begin{algorithmic}[1]
    \STATE Construct the initial Hamitonian $H$ and evaluate
           the exchange energy using $\Phi_{\mathrm{ref}}$;

    \WHILE {Exchange energy is not converged}

    \STATE Solve the linear eigenvalue problem $H[P]\psi_{i} =
    \varepsilon_{i} {\psi}_{i}$ using an iterative eigensolver.

    \STATE Evaluate $\Phi, R_{\Phi}$ according to
    Eq.~\eqref{eqn:Phitransform},~\eqref{eqn:residualC}.

    \STATE Perform the DIIS procedure according to Eq.~\eqref{eqn:Pcp2} to obtain
    the new $\tilde{\Phi}$ which implicitly defines a density matrix
    $P$ via Eq.~\eqref{eqn:PPhi}.

    \STATE Update the Hamiltonian $H[P]$.

    \STATE Compute the exchange energy.

    \ENDWHILE
  \end{algorithmic}
  \label{alg:pcdiis}
\end{algorithm}

The discussion above is applicable when $\Psi$ only contains the
occupied orbitals. When $\Psi$ also involves the unoccupied
orbitals, we use the fact that the density matrix defining the Fock
exchange operator only involves the occupied orbitals, and we only
need to apply the PC-DIIS method to the occupied orbitals. We also
remark that the PC-DIIS method is not yet applicable for finite
temperature calculations with fractionally occupied orbitals, and this
will be our future work.

\section{Wavefunction extrapolation in \textit{Ab initio} molecular
dynamics}\label{sec:aimd}

In \textit{ab initio} molecular dynamics (AIMD) simulation, the
electron density and the Kohn-Sham orbitals between consecutive MD
steps are correlated. Hence one can extrapolate 
electron density density and/or Kohn-Sham orbitals from 
previous MD steps to produce an initial guess for the new
MD step. The simplest strategy is a linear extrapolation procedure.
For instance, in KSDFT calculations with semi-local functionals, let
$\rho(t-\delta t)$ and $\rho(t)$ be the electron density at time
$t-\delta t$ and $t$, respectively. One can perform linear
extrapolation
\begin{equation}
  \rho^{p} = 2 \rho(t) - \rho(t-\delta t),
  \label{}
\end{equation}
and use $\rho^{p}$ as the initial guess for the
electron density at time $t+\delta t$. After the self-consistency is
reached at time $t+\delta t$, we 
obtain the corrected electron density $\rho(t+\delta
t)$. For Hartree-Fock-like equations, the Hamiltonian depends on the
entire density matrix, and it is not sufficient to only extrapolate
the electron density. Since it is prohibitively expensive to
extrapolate the density matrix when a large basis set is used, one
can only perform extrapolation on the Kohn-Sham orbitals. However,
due to the arbitrariness in the choice of the gauge matrix, the
Kohn-Sham orbitals $\Psi(t)$ may depend on the choice of the gauge
in the eigensolver. In this case, $\Psi(t)$ is not even be
continuous with respect to $t$.

Nonetheless, the density matrix $P(t)$ is smooth with respect to
$t$. Our main observation is that if we can choose the
time-dependent gauge-fixing matrix $\Phi_{\mathrm{ref}}(t)$ that is
smooth with respect to $t$, then the corresponding matrix
$P(t)\Phi_{\mathrm{ref}}(t)$ will also be smooth respect to $t$. In
particular, according to Eq.~\eqref{eqn:PPhi}, the information
contained in $P(t)$ and $P(t)\Phi_{\mathrm{ref}}(t)$ are equivalent.
For each given time $t$, the only constraint in the choice of
$\Phi_{\mathrm{ref}}(t)$ is that $P(t)\Phi_{\mathrm{ref}}(t)$ should
have full column rank. Assuming we have already obtained
full rank $\Phi_{\mathrm{ref}}(t-\delta t)$ and $\Phi_{\mathrm{ref}}(t)$,
then we can perform linear extrapolation
\begin{equation}
  \Phi_{\mathrm{ref}}^p = 2 \Phi_{\mathrm{ref}}(t) -
  \Phi_{\mathrm{ref}}(t-\delta t)
  \label{}
\end{equation}
to obtain the predicted gauge-fixing matrix $\Phi_{\mathrm{ref}}^p$. It follows
from Eq.~\eqref{eqn:PPhi} again that the density matrix associated with
$\Phi_{\mathrm{ref}}^p$ is defined by
\begin{equation}
  P^{p} = \Phi_{\mathrm{ref}}^p
  \left[\left(\Phi_{\mathrm{ref}}^p\right)^{*}\Phi_{\mathrm{ref}}^p\right]^{-1}
  \left(\Phi_{\mathrm{ref}}^p\right)^{*}.
  \label{eqn:Ppredict}
\end{equation}
We stress that we never explicitly construct the density matrix
$P^{p}$, but only implicitly use the matrix factors in Eq.~\eqref{eqn:Ppredict} 
to update the electron density and the exchange operator. After self-consistency is reached at $t+\delta t$, we obtain the Kohn-Sham orbitals
denoted by $\Psi(t+\delta t)$. This gives us the corrected 
gauge-fixing matrix at time $t+\delta t$ as
\begin{equation}
  \Phi_{\mathrm{ref}}(t+\delta t) = P(t+\delta t) \Phi_{\mathrm{ref}}^p
  = \Psi(t+\delta t)
  \left(\Psi^{*}(t+\delta t)
  \Phi_{\mathrm{ref}}^p\right),
  \label{}
\end{equation}
which is clearly gauge invariant with respect to
$\Psi(t+\delta t)$. Again due to Eq.~\eqref{eqn:PPhi}, the
$\Phi_{\mathrm{ref}}(t+\delta t)$ and $P(t+\delta t)$
span the same space, and the MD simulation can continue. Our numerical
results indicate that the extrapolation of the density matrix by
means of the gauge-fixing matrix can effectively reduce the number
of SCF iterations in AIMD simulation.

\section{Numerical results} \label{sec:Result}


We demonstrate the accuracy and efficiency of the PC-DIIS method
using the DGDFT (Discontinuous Galerkin Density Functional Theory)
software package.\cite{JCP_231_2140_2012_DGDFT,
JCP_143_124110_2015_DGDFT, PCCP_17_31397_2015_DGDFT,
JCP_145_154101_2016_DGDFT, JCP_335_426_2017_DGDFT} DGDFT is a
massively parallel electronic structure software package designed
for large scale DFT calculations involving up to tens of thousands
of atoms. It includes a self-contained module called PWDFT for
performing planewave based electronic structure calculations (mostly
for benchmarking and validation purposes). We implemented the
PC-DIIS method in PWDFT. We use the Message Passing Interface (MPI)
to handle data communication, and the Hartwigsen-Goedecker-Hutter
(HGH) norm-conserving pseudopotential\cite{PRB_58_3641_1998_HGH}.
All calculations use the HSE06
functional.\cite{JCP_124_219906_2006_HSE06} All calculations are
carried out on the Edison systems at the National Energy Research
Scientific Computing Center (NERSC). Each node consists of two Intel
``Ivy Bridge'' processors with $24$ cores in total and 64 gigabyte
(GB) of memory. Our implementation only uses MPI. The number of
cores is equal to the number of MPI ranks used in the simulation.

\begin{figure}[htbp]
\begin{center}
\includegraphics[width=0.85\textwidth]{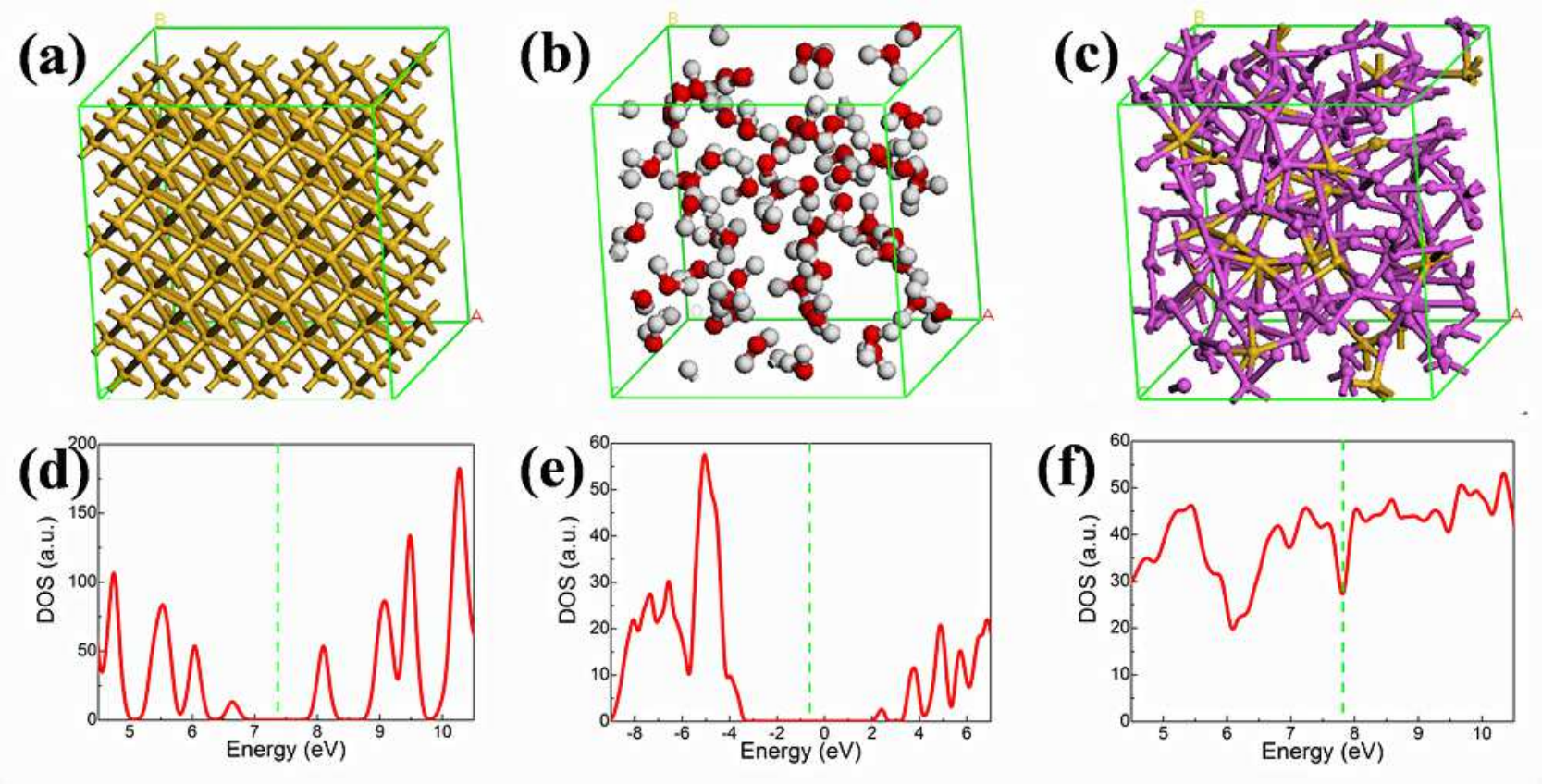}
\end{center}
\caption{(Color online) Atomic structures of (a) Si$_{216}$, (b)
(H$_2$O)$_{64}$ and (c) Al$_{176}$Si$_{24}$. The white, red, yellow
and pink balls denote hydrogen, oxygen, silicon and aluminum atoms,
respectively. The total densities of states (DOS) of semiconducting
(d) Si$_{216}$ and (e) (H$_2$O)$_{64}$, and (f) metallic
Al$_{176}$Si$_{24}$. The Fermi levels are marked by green dotted
lines.} \label{fig:DOS}
\end{figure}

In this section, we demonstrate the performance of the PC-DIIS
method for accelerating hybrid functional calculations by using six
different systems. They consist of four bulk silicon systems (Si$_{64}$,
Si$_{216}$, Si$_{512}$ and Si$_{1000}$),\cite{JCTC_12_2242_2016_ACE}
a bulk water system with $64$ molecules ((H$_2$O)$_{64}$) and a
disordered silicon aluminum alloy system
(Al$_{176}$Si$_{24}$)\cite{JCP_145_154101_2016_DGDFT} as shown in
Figure~\ref{fig:DOS}. Bulk silicon systems (Si$_{64}$, Si$_{216}$,
Si$_{512}$ and Si$_{1000}$) and bulk water system ((H$_2$O)$_{64}$)
are semiconducting with a relatively large energy gap $E_\text{gap}
> 1.0$ eV, and the Al$_{176}$Si$_{24}$ system is metallic with a
small energy gap $E_\text{gap} < 0.1$ eV. The density of states of
these systems are given in Figure~\ref{fig:DOS}. All systems are
closed shell systems, and the number of occupied bands is
$N_\text{band} = N_{e}/2$. In order to compute the energy gap in the
systems, we also include two unoccupied bands in all calculations.
%

\subsection{Accuracy}\label{sec:Accuracy}

We first validate the accuracy of the PC-DIIS method compared to the
two-level nested SCF procedure for hybrid functional calculations.
In both cases we use the adaptively compressed exchange
(ACE)\cite{JCTC_12_2242_2016_ACE} formulation to accelerate
calculations, which reduces the number of times the
exchange operator is applied to Kohn-Sham orbitals without loss of accuracy.
Table~\ref{Accuracy} shows the differences between the energy gaps,
the HF energies, the total energies and the atomic forces computed
by PC-DIIS and a nested two-level SCF procedure respectively. 
The HF and total energy differences as well as the difference in atomic
forces are define by
\[
\Delta{E_\text{HF}} = (E_\text{HF}^\text{PC-DIIS} -
E_\text{HF}^\text{NESTED})/N_{A},
\]
\[
\Delta{E} = (E^\text{PC-DIIS} - E^\text{NESTED})/N_{A},
\]
\[
{\Delta}F = \max_I|F_I^\text{PC-DIIS} - F_I^\text{NESTED}|,
\]
where the superscript NESTED denotes quantities obtained from
a nested two-level SCF procedure, and $N_{A}$ is the total number of atoms 
and $I$ is the atom index.

All calculations start
from initial Kohn-Sham orbitals obtained from converged calculations
using the PBE functional.\cite{PRL_77_3865_1996_PBE} 
The kinetic
energy cutoff $E_{\text{cut}}$ is set to $10$ Hartree for bulk
silicon systems, $40$ Hartree for the Al$_{176}$Si$_{24}$ system,
and $60$ Hartree for water system, respectively. 
%
\begin{table}
\caption{The differences between the HF and total energies and atomic forces computed
by a nested two-level SCF procedures and by the PC-DIIS method.
The unit for HF and total energy difference is Hartree/atom. The unit atomic for
force difference is Hartree/Bohr. Column 2 lists the energy
gap in eV and the energy difference (in eV) is shown in the parenthesis.
}
\label{Accuracy}
\begin{tabular}{ccccc} \\
\hline \hline
Systems  &  $E_\text{gap}$ & ${\Delta}E_\text{HF}$ & ${\Delta}E_\text{tot}$ & ${\Delta}F$   \ \\
\hline
  Si$_{64}$          &    1.484561 (1.10E-07)  &  1.56E-09  &  1.25E-08  &  4.45E-06  \ \\
 Si$_{216}$          &    1.449789 (3.00E-08)  &  4.63E-10  &  4.62E-09  &  7.27E-07  \ \\
 Si$_{512}$          &    1.324900 (8.00E-07)  &  1.95E-09  &  1.95E-08  &  1.56E-05  \ \\
Si$_{1000}$          &    1.289140 (1.00E-08)  &  1.00E-09  &  1.00E-08  &  5.30E-07  \ \\
\hline
(H$_2$O)$_{64}$      &    5.991825 (5.00E-07)  &  2.60E-08  &  1.04E-07  &  9.59E-07  \ \\
\hline
Al$_{176}$Si$_{24}$  &    0.098631 (2.20E-05)  &  5.50E-09  &  2.15E-07  &  4.10E-06  \ \\
\hline \hline
\end{tabular}
\end{table}

Our calculations indicate that the electronic properties obtained from
the PC-DIIS method is fully comparable to that obtained from the
nested two-level SCF procedure, for both semiconducting and metallic
systems. The remaining difference between the two methods is
comparable to the residual error in the SCF iteration, and can
further be reduced with a tighter convergence criterion.


\subsection{Efficiency}\label{sec:Efficiency}

We demonstrate the efficiency of the PC-DIIS method by performing
hybrid DFT calculations for a bulk silicon system with 1000 atoms
($N_\text{band} = 2000$) on 2000 computational cores. In both
PC-DIIS and the nested two-level SCF procedure, we use the ACE-ISDF method\cite{HuLinYang2017}
to accelerate hybrid functional calculations within each SCF
iteration. A direct application of the exchange operator to
$N_\text{band}=2000$ occupied orbitals requires the solution of
$N_\text{band}^2 = 4,000,000$ Poisson-like equations. 
The ISDF method compresses these
equations into $c N_\text{band}$ equations, where $c$ is referred to
as the rank parameter. Following Ref.~\onlinecite{HuLinYang2017}, we
choose the rank parameter $c=6$, which reduces the number of
Poisson-like equations to merely $12,000$. The corresponding error
in the energy and atomic force is below $10^{-3}$ Hartree/atom and
$10^{-3}$ Hartree/Bohr, respectively. For comparison, we also report
the wall clock time for conventional hybrid functional calculations,
which uses the nested two-level method  without the ACE-ISDF
method. The results are summarized in Table~\ref{Efficiency}.
\begin{table}
\caption{A comparison of the computational efficiency
exhibited by PC-DIIS and nested two-level SCF procedures in which 
with and without using ACE-ISDF technique 
on the Si$_{1000}$ system using 2000
computational cores. We report the wall clock time (second) per
outer iteration. For the nested two-level method method, we also report 
the number of inner SCF iterations for each outer SCF iteration.} \label{Efficiency}
\begin{tabular}{ccccccccc} \\ \hline \hline
Outer SCF   & & \multicolumn{2}{c}{PC-DIIS} & &  \multicolumn{4}{c}{NESTED (Two-level SCF procedure)} \ \\
iteration & & \multicolumn{2}{c}{ACE-ISDF} & &  \multicolumn{2}{c}{ACE-ISDF} &  \multicolumn{2}{c}{Conventional}  \ \\
number & & \multicolumn{2}{c}{Time}   & & \#Inner  & Time  &  \#Inner  & Time  \ \\
\hline
1th  & & \multicolumn{2}{c}{38.36}  & & 11 &  360.94  &  5  &  1962.32 \ \\
2nd  & & \multicolumn{2}{c}{39.14}  & & 8  &  145.48  &  4  &  1565.80 \ \\
3rd  & & \multicolumn{2}{c}{39.08}  & & 6  &  114.78  &  4  &  1566.30 \ \\
4th  & & \multicolumn{2}{c}{39.70}  & & 3  &  62.67   &  2  &  831.04  \ \\
5th  & & \multicolumn{2}{c}{38.31}  & & 2  &  48.50   &  1  &  460.96  \ \\
6th  & & \multicolumn{2}{c}{39.01}  & & 1  &  34.82   &  -  &  -  \ \\
\hline \hline
\end{tabular}
\end{table}

\begin{figure}[htbp]
\begin{center}
\includegraphics[width=0.5\textwidth]{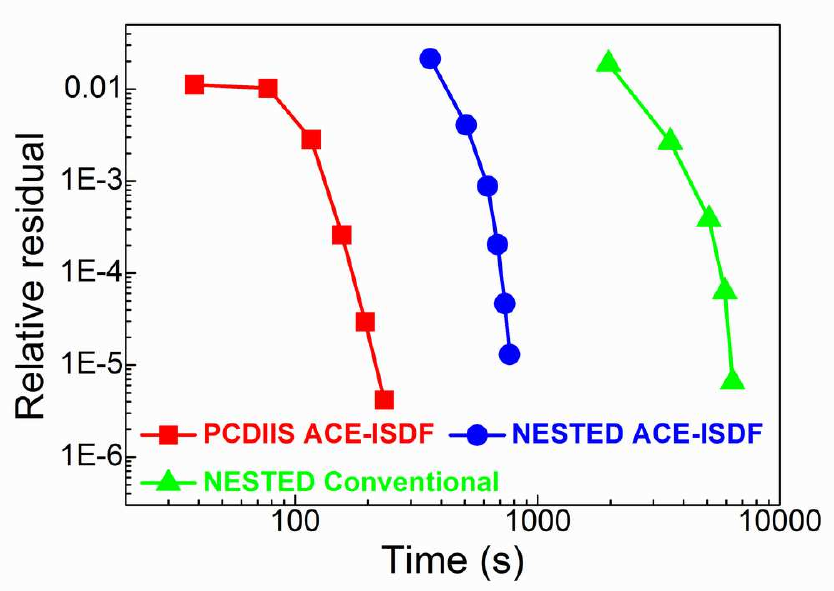}
\end{center}
\caption{(Color online) The relative residual of HF energy with
respect to the wall clock time computed with one-level (PC-DIIS) and
two-level (NESTED) SCF procedures for ACE-ISDF enabled and
conventional hybrid functional calculations on the Si$_{1000}$
system on 2000 cores.} \label{fig:dExx}
\end{figure}

The nested two-level SCF procedure requires
$6$ and $5$ outer SCF iterations to converge regardless whether 
ACE-ISDF is used. A similar number of outer SCF iterations is required 
in the PC-DIIS method also. Hence the total number of iterations from
the PC-DIIS method is very comparable to the number of outer SCF
iterations in the nested two-level SCF method.  However, the nested 
two-level SCF method involves many inner SCF
iterations especially at the beginning stage, which increases the
computational time significantly. Figure~\ref{fig:dExx} shows the
relation between the residual of the HF energy in hybrid
functional calculations with respect to the wall clock time, plotted
on a logarithmic scale. The PC-DIIS method only takes $233.6$ s to
converge the entire simulation, which is significantly faster than
the two-level SCF procedure with ACE-ISDF ($767.19$ s) and without ACE-ISDF
($6847.28$ s), respectively.

\subsection{AIMD}\label{sec:AIMD}

%
%

We demonstrate the performance of the PC-DIIS method in the context of
AIMD simulation for the Si$_{64}$ system under the NVE ensemble, and the
(H$_2$O)$_{64}$ system under the NVT ensemble, respectively.

\begin{figure}[htbp]
\begin{center}
\includegraphics[width=0.85\textwidth]{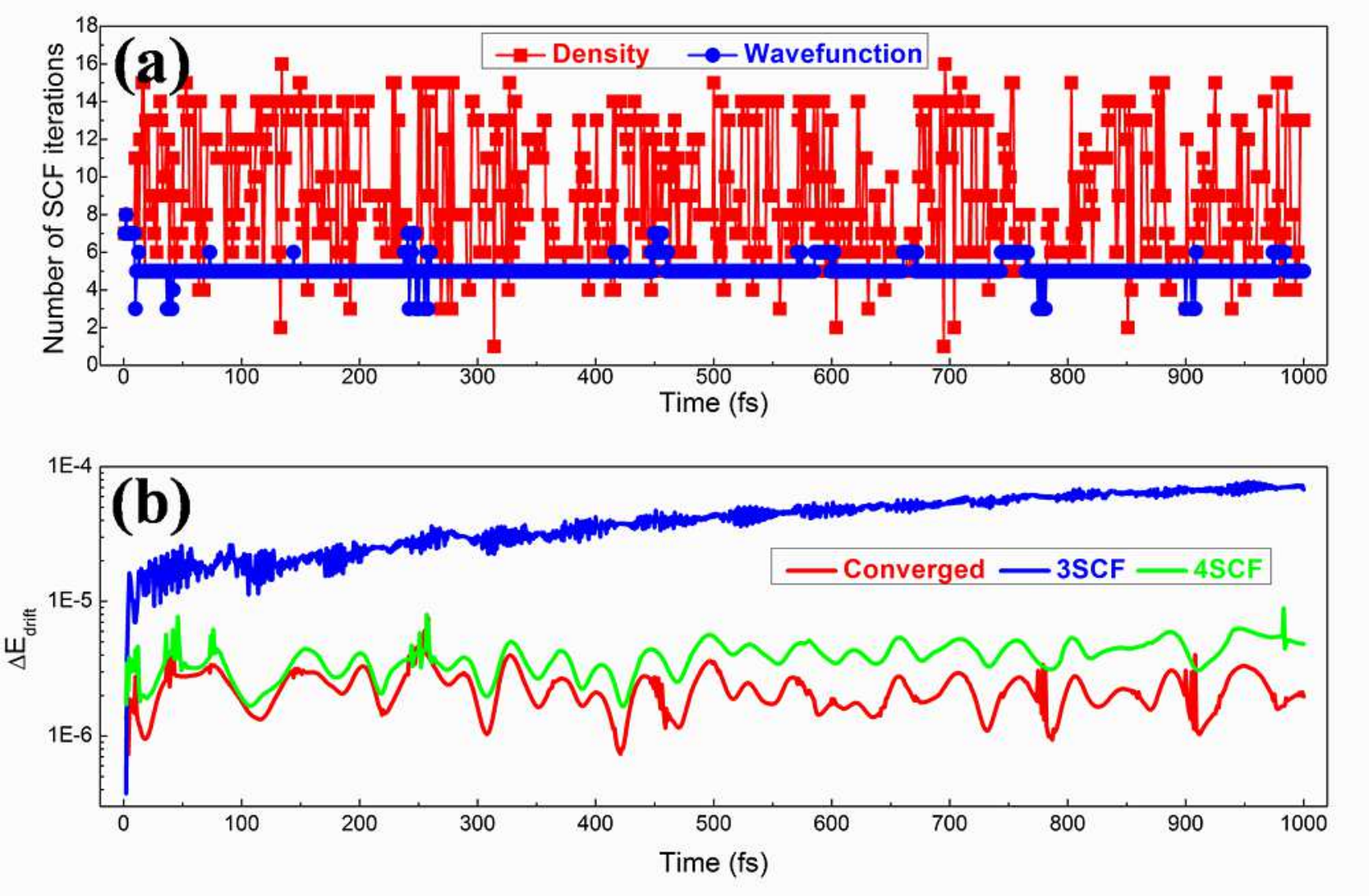}
\end{center}
\caption{(Color online) Comparison of AIMD simulations by using
one-level (PC-DIIS) and two-level (NESTED) SCF procedures for ACE
enabled hybrid functional calculations on the Si$_{64}$ system. (a)
The number of SCF iterations used per MD step using linear
extrapolation of the density (``density''), and linear extrapolation
of the gauge-fixing matrix (``wavefunction''), respectively. (b) The
relatively energy drift by comparing different choices of maximum
number of SCF iterations for the PC-DIIS method per MD step.}
\label{fig:Si64MD}
\end{figure}

Figure~\ref{fig:Si64MD} (a) shows the number of SCF iterations along
AIMD trajectory for the Si$_{64}$ system, using linear extrapolation
of the electron density (``density''), and linear extrapolation of
the gauge-fixing matrix (``wavefunction''), respectively. In the
former case, the initial guess for the density matrix at time
$t+\delta t$ is given by the output density matrix at time $t$. In
the latter case, the extrapolated gauge-fixing matrix
$\Psi_{\mathrm{ref}}^{p}$ provides the initial guess for both the
density matrix and the density in a consistent fashion at time
$t+\delta t$. The convergence criterion is 10$^{-6}$, and the time
step is $1.0$ femtosecond (fs). We find that the density
extrapolation requires on average $14$ iterations per MD step, while
the wavefunction extrapolation only requires $5$ iterations per MD
step, respectively. We also observe that the variance of the number
of iterations in the wavefunction extrapolation is significantly
smaller.

In AIMD simulation under the NVE ensemble, the total energy is
conserved, and hence the error of the numerical scheme can also be
measured in terms of the relative energy drift, defined as $\Delta
E_{\mathrm{drift}}(t+\delta t)) = (E_{\mathrm{tot}}(t+\delta t)) -
E_{\mathrm{tot}}(t))/E_{\mathrm{tot}}(t)$. Figure~\ref{fig:Si64MD}
(b) reports the relative drift along the MD trajectory, with
different choices of maximum number of SCF iterations per MD step in
the PC-DIIS method. The linear extrapolation of the gauge-fixing
matrix is used as the initial guess. We find that using a maximum of
3 SCF iterations already lead to small but noticeable linear drift
in the total energy, while the energy drift becomes significantly
smaller and stable when a maximum number of $4$ SCF iterations is
used per MD step.

We also apply the PC-DIIS method to perform AIMD simulations on
liquid water system (H$_2$O)$_{64}$ at $T$ = $295$ K. We use a
single level Nose-Hoover thermostat\cite{JCP_81_511_1984_Nose,
PRA_31_1695_1985_Hoover}, and the choice of mass of the Nose-Hoover
thermostat is $85000$ au. The MD time step size is $1.0$ femtosecond
(fs). After equilibrating the system starting from a
prepared initial guess,\cite{JCP_141_084502_2014} we perform the
simulation for 2.0 ps to sample the radial distribution function.
The average number of converged SCF iterations per MD step is 10. We
compare the results from HSE06 and PBE functionals. In both cases
the Van der Waals (VdW) interaction is modeled at the level of the
DFT-D2 method.\cite{JCC_27_1787_2006_Grimme} We also benchmark our
result with the experimental measurement from X-ray diffraction
technique.\cite{JCP_138_074506_2013} Although the second shell
structure is not yet converged due to the relative short simulation
length, the structure from the first shell is already clear. We
observe that the PBE functional leads to over-structured radial
distribution function, which is reduced by the HSE06 functional.
This behavior is in quantitative agreement with previous hybrid
functional DFT calculations,\cite{JCP_141_084502_2014} where the
remaining difference with respect to the experimental result can be
to a large extent attributed to the nuclei quantum
effects.\cite{MorroneCar2008}

\begin{figure}[htbp]
\begin{center}
\includegraphics[width=0.5\textwidth]{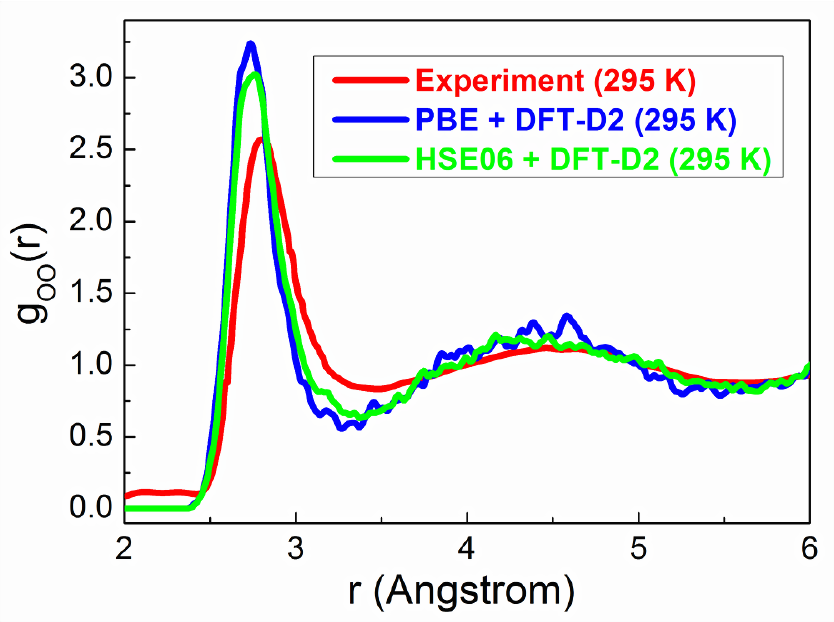}
\end{center}
\caption{(Color online) The oxygen-oxygen radial distribution
functions $g_\text{OO}$($r$) of liquid water system (H$_2$O)$_{64}$
at $T$ = $295$ K obtained from AIMD simulations with semilocal
GGA-PBE and hybrid HSE06 functional calculations with van der Waals
correction (DFT-D2), compared to experimental measurement.}
\label{fig:H2O64gOO}
\end{figure}


\section{Conclusion} \label{sec:conclusion}

We developed the projected commutator-DIIS (PC-DIIS) method for
accelerating the self-consistent field (SCF) iteration in Kohn-Sham
density functional theory calculations with hybrid
exchange-correlation functionals. The PC-DIIS method is particularly
well suited in the context of a large basis set such as the planewave
basis set, where it is prohibitively expensive to even store the
density matrix and the Hamiltonian matrix. The key idea is to
project both the density matrix and the commutator matrix to an
auxiliary matrix called the gauge-fixing matrix
$\Phi_{\mathrm{ref}}$. Then we can extrapolate the projected
matrices that are gauge-invariant, and share the same dimension as
the Kohn-Sham orbitals. This procedure also implicitly updates the
entire density matrix and hence the Hamiltonian matrix in SCF
iterations. Compared to the commonly used two-level nested SCF
structure used for hybrid functional calculations, the PC-DIIS
method only involves one SCF loop, without sacrificing the accuracy
or the convergence rate. In the context of \textit{ab initio}
molecular dynamics (AIMD) simulation, the gauge-fixing matrix
further provides an efficient and gauge-invariant way for implicit
extrapolation of the density matrix. Numerical results indicate that
the new extrapolation scheme significantly reduces the number of SCF
iterations compared to the commonly used strategy of only
extrapolating the electron density.

The PC-DIIS method can be directly extended along several
directions. The two-level nested SCF structure is not only used in
hybrid functional calculations, but also other contexts such as the
DFT+U calculations.\cite{LiechtensteinAnisimovZaanen1995} Hence the
PC-DIIS method can be potentially useful for accelerating such
calculations. For AIMD simulation, the new extrapolation scheme of
the gauge-fixing matrix can be combined with time-reversible
integrators such as the extended Lagrangian Born-Oppenhemier
molecular dynamics (XL-BOMD)
method\cite{PRL_100_123004_2008_Niklasson_XLBOMO} and the almost
stable predictor-corrector (ASPC)
method\cite{JCC_25_335_2004_Kolafa_ASPC} to further reduce the
number of SCF iterations. The freedom in the choice of
$\Phi_{\mathrm{ref}}$ allows one to consider choosing this matrix to
be a sparse matrix, so that the PC-DIIS method can be used in the
context of linear scaling methods. Finally, the PC-DIIS method is
not yet applicable to systems under finite temperature with
fractionally occupied orbitals. We will explore these directions in
the near future.

%
%
%
%
%

\section{Acknowledgments}

This work was partly supported by the National Science Foundation
under grant DMS-1652330 (L. L.), by the Scientific Discovery through
Advanced Computing (SciDAC) program funded by U.S.~Department of
Energy, Office of Science, Advanced Scientific Computing Research
and Basic Energy Sciences (W. H., L. L. and C. Y.), and by the
Center for Applied Mathematics for Energy Research Applications
(CAMERA) (L. L. and C. Y.). The authors thank the National Energy
Research Scientific Computing (NERSC) center and the Berkeley
Research Computing (BRC) program at the University of California,
Berkeley for making computational resources available. We thank
Mohan Chen for sharing the information for the (H$_2$O)$_{64}$
system, and Jarrod McClean for discussions related to the C-DIIS
method.


\providecommand{\latin}[1]{#1}
\providecommand*\mcitethebibliography{\thebibliography}
\csname @ifundefined\endcsname{endmcitethebibliography}
  {\let\endmcitethebibliography\endthebibliography}{}

\end{document}